\documentclass{emulateapj}

\newcommand{\be}{\begin{eqnarray}}
\newcommand{\ee}{\end{eqnarray}}
\newcommand{\beq}{\begin{equation}}
\newcommand{\eeq}{\end{equation}}
\def\simless{\mathbin{\lower 3pt\hbox
      {$\rlap{\raise 5pt\hbox{$\char'074$}}\mathchar"7218$}}}
\def\simgreat{\mathbin{\lower 3pt\hbox
      {$\rlap{\raise 5pt\hbox{$\char'076$}}\mathchar"7218$}}} 

\renewcommand{\vec}[1]{\mbox{\boldmath $\displaystyle #1$}}
\newcommand{\grad}{{\mbox{\boldmath $\nabla$}}}

\begin{document}

\title{ Quasi-Periodic Oscillations from Magnetorotational Turbulence}

\author{Phil Arras\altaffilmark{1}, Omer Blaes\altaffilmark{2},
Neal J. Turner\altaffilmark{3} }

\altaffiltext{1}{Kavli Institute for Theoretical Physics, Kohn Hall,
University of California, Santa Barbara, CA 93106;
arras@kitp.ucsb.edu}
\altaffiltext{2}
{Department of Physics, Broida Hall, University of California, Santa
Barbara, CA 93106; blaes@physics.ucsb.edu}
\altaffiltext{3}{Jet Propulsion Laboratory, MS 169-506, California
Institute of  Technology, Pasadena, CA 91109; neal.turner@nasa.jpl.gov}

\begin{abstract}

Quasi-periodic oscillations (QPOs) in the X-ray lightcurves of accreting
neutron star and black hole binaries have been widely interpreted as
being due to standing wave modes in accretion disks. These disks are
thought to be highly turbulent due to the magnetorotational instability
(MRI). We study wave excitation by MRI turbulence in the shearing
box geometry. We demonstrate that axisymmetric sound waves and radial
epicyclic motions driven by MRI turbulence give rise to narrow, distinct
peaks in the temporal power spectrum. Inertial waves, on the other hand,
do not give rise to distinct peaks which rise significantly above the
continuum noise spectrum set by MRI turbulence, even when the fluid
motions are projected onto the eigenfunctions of the modes. This is 
a serious problem for QPO models based on inertial waves.

\end{abstract}


\section{ Introduction }

Millisecond variability has been observed for some time with the Rossi
X-Ray Timing Explorer in both neutron star and black hole binaries
(e.g. McClintock \& Remillard 2004, van der Klis 2004).  In addition
to a broad continuum in the power spectrum, QPOs are observed. These
QPOs have been modeled as hydrodynamic waves in geometrically thin
accretion disks (for reviews see e.g. Wagoner 1999 and Kato 2001),
or ``torus-like'' flows with significant radial pressure gradients
\citep{2004A&A...427..251G,2003MNRAS.344L..37R, 2004ApJ...603L..89K,
2004ApJ...603L..93L,BAF}.

Turbulence seeded by the magnetorotational instability (MRI;
\citealt{bal91}) is widely believed to provide the stresses responsible
for angular momentum transport and accretion.  This raises a series
of questions. Do waves idealized as hydrodynamic perturbations on
a laminar background still exist in turbulent, magnetized accretion
flows? Does the time-averaged flow act as a resonant cavity supporting
standing waves? What is the steady-state wave amplitude due to turbulent
excitation? The simplest and most controlled geometry to explore
these questions is that of the shearing box (e.g. \citealt{haw95}).
One published attempt has been made to look for discrete mode frequencies
in a shearing box MRI simulation \citep{2005AN....326..787B}, with
negative results (see figure 7 of that paper). Analytic estimates for 
turbulent excitation of waves in accretion disks have been
reported in \citet{1993ApJ...418..187N, 1995MNRAS.274...37N}.

We revisit this problem, finding that MRI turbulence does indeed
excite distinct peaks in the power spectrum, and that these peaks can
be identified with certain classes of hydrodynamic modes, specifically
axisymmetric acoustic waves and a global axisymmetric epicyclic oscillation.
Hydrodynamic inertial wave modes, however, are {\it not} clearly detected,
a serious problem for QPO models that are based on these modes
(``$g$-modes'' in diskoseismology parlance).

The plan of the paper is as follows. In section \ref{sec:sbox} we review the
shearing box and compute the linear wave frequencies. We present power
spectra from numerical simulations in section \ref{sec:power}, and discuss
the results.  We present our conclusions and discuss their relevance
to models for QPOs in section \ref{sec:conclusions}.

\section{ The shearing box }
\label{sec:sbox}

We solve the magnetohydrodynamic (MHD) equations as applied
to an isothermal, differentially rotating, magnetized accretion
disk, in the unstratified shearing box approximation:
\be
 \frac{\partial \vec{v}}{\partial t}  +  \vec{v} \cdot \grad \vec{v} &+&
2 \vec{\Omega} \times \vec{v}  = 
 - \frac{1}{\rho} \grad \left( P
 +   \frac{B^2}{8\pi} \right)  \\
 &+&  \frac{\vec{B} \cdot \grad \vec{B} }{4\pi \rho}
 +  2q\Omega^2 x \vec{e}_x
\\
\frac{\partial \rho}{\partial t} & + &  \grad \cdot \left( \rho \vec{v} \right)
 =  0
 \\
\frac{\partial \vec{B} }{\partial t} & = &  \grad \times \left( \vec{v}
\times \vec{B} \right)
\\
P & = & c_s^2 \rho.
\ee
Here $\rho$, $P$, $\vec{v}$, and $\vec{B}$ are the mass density,
gas pressure, velocity and magnetic field. The sound speed $c_s$
is constant. The simulation domain represents a small region of the
disk near the midplane orbiting with frequency $\Omega$. The
non-inertial reference frame of the orbit is taken into account by
including a Coriolis force, as well as ``tidal" forces due to the
difference of gravitational and centrifugal forces, represented by
the term $2q\Omega^2 x \vec{e}_x$, where $q=-d\ln \Omega/d\ln R$
is the shear parameter.  Cartesian coordinates are used, with the
radial, azimuthal, and vertical coordinates labeled $x$, $y$, $z$,
respectively. The azimuthal and vertical boundaries are periodic,
and the radial boundaries are shearing-periodic (fluid passing
through one radial boundary appears on the other at an azimuth that
varies in time according to the difference in orbital speed across
the box).

An initially weak magnetic field will drive turbulence in the shearing box
due to the MRI, which is a nearly
incompressible instability.  A second branch of nearly incompressible
perturbations which we call ``inertial waves'' also exists, at least on a
laminar background flow.
In the limit of zero magnetic field, the specific angular momentum
gradient is the restoring force for these inertial waves. To understand the
two solutions with a simple example,  
linearize about a time-independent background with constant density $\rho$
and pressure $P$, constant 
magnetic field $\vec{B}=B_y \vec{e}_y + B_z \vec{e}_z$, and velocity 
$\vec{v}=-q\Omega x \vec{e}_y$. For incompressible, axisymmetric waves
with space-time dependence
$\exp(ik_x x + ik_z z - i\omega t)$, the dispersion relation is
\citep{bal91}
\beq
\omega^2  = k_z^2 v_{{\rm A}z}^2 + \frac{1}{2} \kappa^2 \frac{k_z^2}{k^2}
\pm \frac{1}{2} \left[ \left(\kappa \frac{k_z}{k} \right)^4
+ 16 k_z^2 v_{{\rm A}z}^2 \Omega^2 \frac{k_z^2}{k^2} \right]^{1/2},
\label{eq:incomp}
\eeq
where the positive sign gives inertial waves and the negative sign the
MRI. Here $v_{{\rm A}z}=B_z/(4\pi \rho)^{1/2}$ is the vertical Alfven speed and
$\kappa=[2(2-q)]^{1/2}\Omega$ is the fluid epicyclic frequency.\footnote{
  Not to be confused with the epicyclic frequency of a particle in a
  gravitational field. }
In the zero field limit, incompressible inertial waves
have the dispersion relation $\omega^2=\kappa^2 k_z^2/k^2$.

Although inertial modes modified by magnetic tension exist for the
constant background chosen above, once the medium becomes turbulent,
their possible existence as normal modes of oscillation becomes highly
questionable. Low order inertial waves have frequencies of the same
order as the energy-bearing MRI ``eddies'', and hence are susceptible to
nonlinear interactions with the MRI turbulence.

The simulations described here have zero net magnetic flux in each
direction. Furthermore, the time and space-averaged magnetic pressure
is small. We attempt to understand waves in such a turbulent background
by again linearizing about a time-independent background with constant
density and pressure, velocity $\vec{v}=-q\Omega x \vec{e}_y$, but now
including compressibility and neglecting magnetic fields.  In this case,
axisymmetric inertial-acoustic waves satisfy the dispersion relation
\beq
\omega^2  =  \frac{1}{2} \left( \kappa^2 + c_s^2 k^2 \right) \pm
\frac{1}{2} \left[ \left(\kappa^2+c_s^2 k^2 \right)^2 
- 4\kappa^2 c_s^2 k_z^2 \right]^{1/2}.
\label{eq:disprel}
\eeq
For zero wavenumber one recovers the fluid epicyclic mode
$\omega^2=\kappa^2$. In the limit $\kappa^2 \ll c_s^2 k^2$, which is
a good approximation even for the
lowest wavenumbers in our simulations, inertial
waves again have the dispersion relation $\omega^2=\kappa^2
k_z^2/k^2$, while acoustic waves satisfy $\omega^2=c_s^2 k^2$.
In a shearing box of dimensions $(L_x,L_y,L_z)$, these waves will occur
at discrete frequencies because the wavenumbers will be quantized according
to
\beq
k_x={2\pi n_x\over L_x}\,\,\,\,\,{\rm and}\,\,\,\,\,k_z={2\pi n_z\over L_z},
\eeq
where $n_x$ and $n_z$ are integers.

Because of the shearing boundary conditions, non-axisymmetric waves in
the shearing box have an amplitude and frequency which change with time
\citep{gol65}, and therefore do not exist as normal modes. The radial
wavenumber  changes with time as $k_x(t)=k_x'+q\Omega (t-t_0) k_y'$,
where $k_x'$ and $k_y'$ are the wavenumbers with respect to comoving
coordinates $[x,y'=y+q\Omega x (t-t_0)]$, and $t_0$ is the time of minimum
$k_x(t)$. When $k_x(t) \sim k_x'$, non-axisymmetric waves have similar
frequencies to axisymmetric waves. For $k_x(t) \gg k_x'$, an approximate
WKB dispersion relation for the time-dependent frequency is (see eq. 72 in
\citealt{gol65}) $\omega(t)\simeq c_s k_y' q\Omega t$, showing  that the
frequency changes on the shear timescale of the disk. As $k_x(t)$ becomes
large, eventually the wave will damp away. If the wave has lifetime $T$,
then we expect it to give rise to a broad peak in the power spectrum of
width $\sim ck_y'q\Omega T$ around the corresponding axisymmetric mode.

\section{ Power spectra } 
\label{sec:power}

\begin{figure*}
\epsscale{0.9}
\plotone{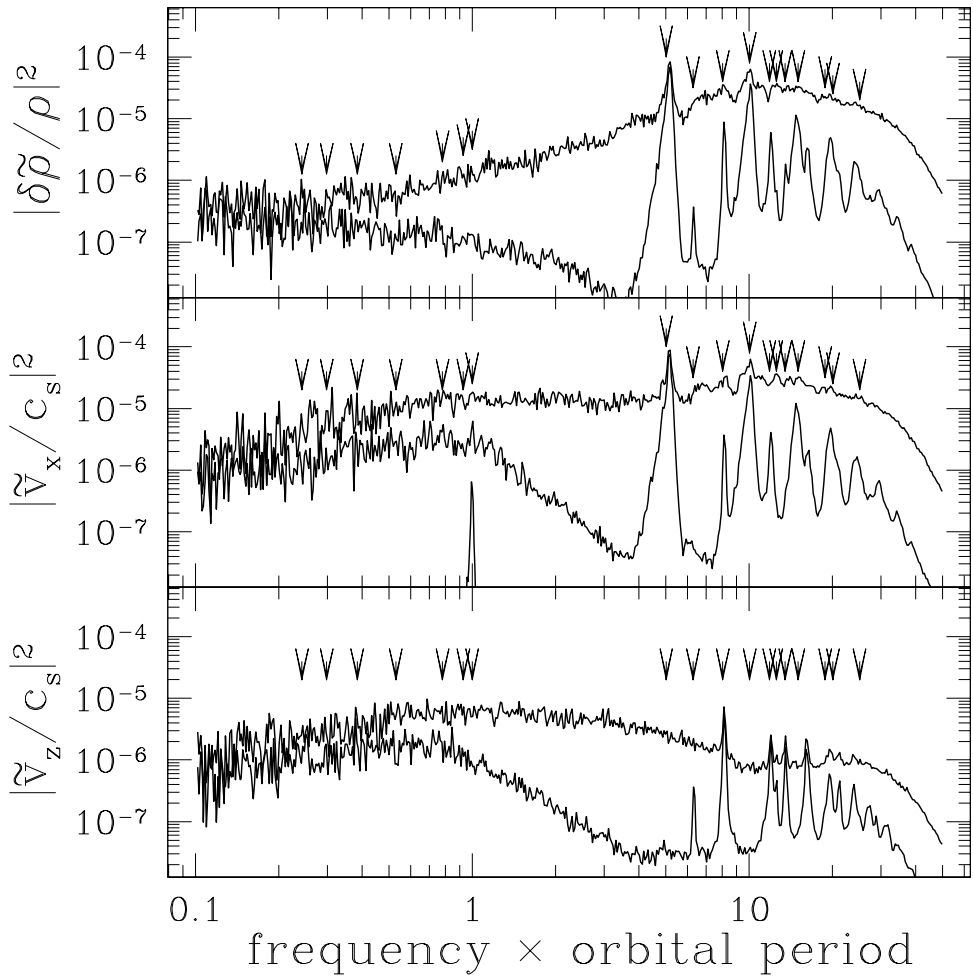}
\caption{``Position space" power spectrum for density
  (upper), radial velocity (middle), and vertical velocity
  fluctuations (lower). In each plot, the upper curve is for the
  non-axisymmetric data and the lower curve is for the axisymmetric
  data. The sharp peak at the orbital frequency in the middle plot is
  the box averaged data. The frequency is in units of cycles per
  orbital period. Power has been rebinned into logarithmically spaced
  frequency bins, which effectively multiplies the power spectrum by
  one power of frequency. Axisymmetric hydrodynamic normal mode frequencies
  from eq.\ (\ref{eq:disprel}) are shown as arrows.  The epicyclic mode
  is at the orbital frequency (unity). Inertial modes and sound waves
  should lie below and above the epicyclic frequency, respectively. From
  right to left below the epicyclic frequency, arrows show the expected
  axisymmetric inertial mode frequencies for $(n_x,n_z)=(1,2), (1,1),
  (2,1), (3,1), (4,1)$ and $(5,1)$. From left to right above the epicyclic
  frequency, the axisymmetric acoustic waves have $(n_x,n_z)=(1,0),(0,1),
  (1,1),(2,0),(2,1),(0,2),(1,2),(3,0),(0,3),(4,0)$, and $(5,0)$,
  respectively. } 
\label{fig:power}
\end{figure*}

Our simulations use box size $(L_x,L_y,L_z)=(1.25,4,1)$ with
$(N_x,N_y,N_z)=(40,64,32)$ grid points in each direction, respectively.
The shear parameter $q=3/2$, corresponding to Keplerian rotation.
The net magnetic flux in each direction is zero and the initial weak
field has a spatial dependence given by ${\bf B}=B_0\cos(2\pi x/L_x){\bf
e}_y+B_0\sin(2\pi x/L_x){\bf e}_z$. The initial magnetic pressure is 400
times less than the gas pressure. The sound speed, orbital frequency,
and average density have the constant values $c_s=\Omega=10^{-3}$
and $\rho=1$. The fiducial simulation is run for 1000 orbits. Averaged
over time and the box, the total magnetic pressure in the turbulence
that results is $0.65\%$ of the average gas pressure.  However, there
is significant variance, with small regions of local magnetic pressure
reaching $10\%$ of the gas pressure.

Figure \ref{fig:power} shows ``position space" power spectra for density,
radial velocity, and vertical velocity fluctuations.
Three spectra are plotted for each. Data
recorded at a specific point $(x,y,z)$ in the box contain fully
non-axisymmetric motions. Fluid motions with nodes along the $y$-direction
contain most of the power in MRI turbulence, hence we also record time
series for azimuthally averaged data at the same $(x,z)$ and box-averaged
data in order to increase signal to noise for oscillatory modes.
We interpolated the data to uniform spacing in time,  padded the array
to a multiple of two, multiplied by the Bartlett window function
to reduce spectral leakage \citep{1992nrfa.book.....P}, and used
a Discrete Fast Fourier Transform to produce the power spectrum.
The power spectra are normalized to the root-mean-square of the time
series. To produce figure \ref{fig:power} we binned the power into
512 logarithmically spaced frequency bins in the frequency range
shown. This has the effect of multiplying the power spectrum by
a factor of $\omega$. The frequency $\omega/2\pi$ is expressed in
units of inverse orbital time.  Arrows show the position of low order
acoustic and inertial waves using eq.\ (\ref{eq:disprel}). While all
three panels show curves for non-axisymmetric and axisymmetric data,
the box-averaged power is only visible in the plot for $v_x$, appearing
as a sharp peak at $\omega=\kappa=\Omega$, the epicyclic frequency.
This epicyclic mode has been reported in previous shearing box simulations
\citep{haw95}. Numerical damping causes the sharp decline in power at
high frequencies. In runs at higher resolution, this dropoff was pushed
to higher frequencies, as expected.

The most striking feature in figure \ref{fig:power} is the axisymmetric
sound waves, which appear as sharply defined peaks. The power in
these peaks approaches the broad continuum due to non-axisymmetric
MRI turbulence, and clearly rises above the continuum for at least two
of the peaks. Axisymmetric sound waves with $n_z=0$  have much larger
density and radial velocity fluctuations than those with $n_x=0$. This
is consistent with the larger continuum seen in $v_x$ relative to
$v_z$. A number of previous investigations (e.g. \citealt{sto96,gar05})
have noted non-axisymmetric sound waves in snapshots of the density
profile. These non-axisymmetric sound waves also have $n_z=0$ and $n_x
\neq 0$, consistent with our findings for axisymmetric waves.

There are no significant peaks observed for inertial modes in figure
\ref{fig:power}. However, this is not proof of their non-existence.
The epicyclic mode was buried in the noise for non-axisymmetric
and axisymmetric data, and only became visible by box-averaging. Moreover,
the {\it observable} luminosity variation decreases strongly as wavelength
decreases, due to averaging of hot and cold spots on the disk. Hence
large lengthscale, small power inertial modes may {\it in principle} be
observable as compared to short lengthscale, large power MRI eddies. We
now show that this is in fact not the case.

\begin{figure}
\epsscale{1.2}
\plotone{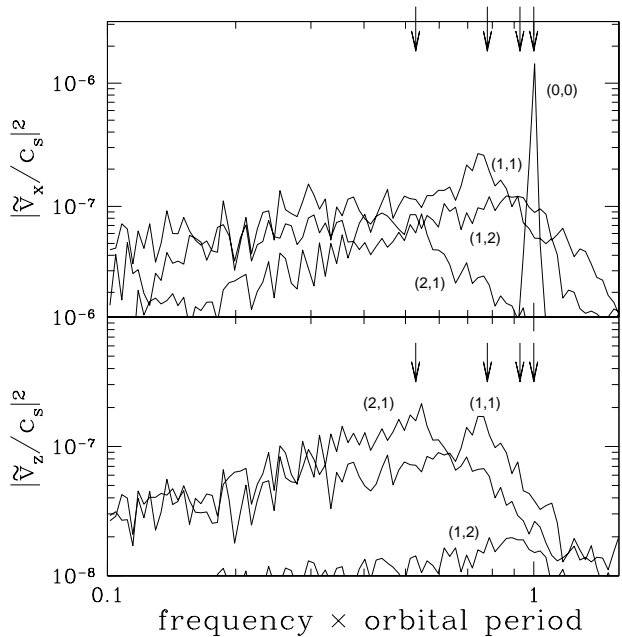}
\caption{ ``Momentum space" power spectrum for axisymmetric
  radial velocity (upper), and vertical velocity (lower) fluctuations.
  Each curve is labelled by $(n_x,n_z)$ used for the spatial Fourier transform.
  Power has been rebinned into logarithmically spaced
  frequency bins, which effectively multiplies the power spectrum by
  one power of frequency. Axisymmetric hydrodynamic inertial mode frequencies
  from eq.\ (\ref{eq:disprel}) are shown as arrows with $(n_x,n_z)=(0,0),
  (1,2), (1,1), (2,1)$ from right to left.}
\label{fig:ftpower}
\end{figure}

Figure \ref{fig:ftpower} shows the ``momentum space" power spectrum.
Time series for azimuthally averaged $v_x$ and $v_z$ were computed by
taking spatial transforms in the $x$ and $z$ directions, for $n_x$ and
$n_z$ nodes, respectively. Hence, turbulent eddies and inertial modes
are on the same footing in this plot, as {\it we compare their power at
the same lengthscale}. The temporal power spectrum for the resulting time
series was computed as in figure \ref{fig:power} for 128 frequency bins in
the range shown. The normalization of the curves is such that the sum over
all $n_x$, $n_z$, and frequency bins gives the root-mean-square of the
time series (compare to axisymmetric curves in figure \ref{fig:power}).

The epicyclic mode is again clearly visible in figure \ref{fig:ftpower}.
The inertial modes are qualitatively different. Two of the lowest
order modes show broak peaks of width $\delta \omega \sim \omega$ at
the expected frequencies. Without the analytic mode frequencies as
a guide, it would be hard to recognize these peaks as significant. 
Higher order modes (not plotted here) show even
less evidence of peaks at the expected frequencies. The lack of clear
peaks for inertial modes suggests that they are strongly affected by MRI
turbulent stresses. We note that although the epicyclic mode has frequency
similar to inertial modes, it is immune to magnetic tension forces as
it is nodeless in the shearing box, and hence is less susceptible to
turbulent buffeting.

Overall, we find that axisymmetrizing or box-averaging the data
from simulations greatly enhances the signal from oscillation modes,
a technique easily adapted to global simulations. In addition, long
integration times will increase the signal to noise of the nearly
sinusoidal waves relative to the stochastic background from turbulence. It
is not immediately obvious, however, why we clearly see peaks in our
non-axisymmetric data whereas \cite{2005AN....326..787B} did not. Perhaps
this is because his simulation was stratified in the vertical direction.

\section{Conclusions} 
\label{sec:conclusions}

We find that turbulence driven by the MRI naturally excites radial
epicyclic motion and axisymmetric sound waves, but there is scant
evidence of inertial waves. This may imply that inertial modes
are strongly affected by MRI turbulent stresses, and cease to exist
as high quality oscillators. We note that the zero net magnetic flux
simulations presented here give rise to the {\it weakest} MRI turbulence
possible. Inertial waves could be even more strongly affected in
simulations with net magnetic flux.

Inertial waves, called ``$g$-modes'' in the diskoseismology literature,
have been proposed as the origin of some observed QPOs in X-ray binaries
\citep{1999PhR...311..259W, 2001PASJ...53....1K}. However, our results
imply that all but the longest wavelength of these hydrodynamic modes
are destroyed by the turbulence. Even the longest wavelength modes are
substantially altered, perhaps to the point of becoming unobservable. This
presents a serious problem for models invoking $g$-modes to explain
observed QPOs.

We are not clearly able to pick out non-axisymmetric, shearing
waves in the analysis done so far. There are broadened peaks in
the non-axisymmetric data around the frequencies of axisymmetric
waves, but this is not clear confirmation. Furthermore, ``real''
accretion disks may in fact admit standing nonaxisymmetric waves
with constant pattern speeds. There is substantial global simulation
data which demonstrates this fact, at least for hydrodynamic disks
(e.g. \citealt{1988ApJ...326..277B}).  The shearing box cannot capture
these modes, because the shearing box boundary conditions preclude
their existence.

In summary, this paper demonstrates that QPOs can be found in simulations
of MRI turbulence, at least in a shearing box geometry.  The well-defined
boundary conditions of this geometry help ensure the existence of modes,
but it is not clear that such well-defined boundaries will exist in real
accretion disks.  Global MRI simulations could in principle determine
self-consistently whether wave cavities exist and contain trapped modes.
Traveling acoustic waves have certainly been observed in such simulations
(e.g.  \citealt{dev03}), but no evidence for modes at discrete frequencies
has yet been reported.  We encourage extensions of our analysis techniques
to these simulations.

\acknowledgements

We thank Steve Balbus, Shane Davis, Mike Nowak and Aristotle Socrates for useful
discussions. This research was supported by the National Science Foundation
under grant nos. PHY99-07949 and AST 03-07657.

\end{document}